\documentclass[conference]{IEEEtran}
\IEEEoverridecommandlockouts
\usepackage{amsmath,amsfonts}
\usepackage{algorithmic}
\usepackage{algorithm}
\usepackage{array}
\usepackage{caption}
\usepackage{subfigure}
\usepackage{textcomp,color}
\usepackage{stfloats}
\usepackage{url}
\usepackage{verbatim}
\usepackage{graphicx}
\usepackage{cite}
\usepackage{amsthm}
\usepackage{hyperref}
\newtheorem{theorem}{Theorem}
\theoremstyle{definition}
\newtheorem{definition}{Definition}
\newtheorem{lemma}{Lemma}
\newtheorem{assumption}{Assumption}
\newtheorem{remark}{Remark}

\DeclareMathOperator*{\argmax}{arg\,max}
\def\BibTeX{{\rm B\kern-.05em{\sc i\kern-.025em b}\kern-.08em
    T\kern-.1667em\lower.7ex\hbox{E}\kern-.125emX}}
\begin{document}

\title{Content Provider Contributions to Capacity Expansion of a Neutral ISP: Effect of Private Option\\
}

\author{\IEEEauthorblockN{Pranay Agarwal}
\IEEEauthorblockA{
BITS Pilani Hyderabad Campus\\ 
Telangana, India - 500078\\
\texttt{agarwalpranay09@gmail.com}}
\and
\IEEEauthorblockN{D. Manjunath}
\IEEEauthorblockA{
IIT Bombay\\
Mumbai, India - 400076 \\
\texttt{dmanju@ee.iitb.ac.in}}
}
\maketitle

\begin{abstract}
Increasing content consumption by users and the expectation of a better Internet experience requires Internet service providers (ISPs) to expand the capacity of the access network continually. The ISPs have been demanding the participation of the content providers (CPs) in sharing the cost of upgrading the infrastructure. From CPs' perspective, investing in the ISP infrastructure, termed as \emph{public investment}, seems rational as it will boost their profit. However, the CPs can alternatively invest in making content delivery more efficient, termed as \emph{private investment}, as it also boosts their profit.
Thus, in this work, we investigate this trade-off between public and private investment of the CPs for a net-neutral ISP. Specifically, we consider centralized decision and non-cooperative forms of interaction between CPs and an ISP and determine the optimum public and private investments of the CPs for each model.  In the non-cooperative interaction, we find that at most one CP contributes to the public infrastructure, whereas all invest in their private infrastructure.  
\end{abstract}
\IEEEpeerreviewmaketitle

\section{Introduction}
Internet users today interact with two complementary service providers---content providers (CPs) that provide the content and access providers or ISPs that provide the connectivity. It is easy to see that content and access are indeed complementary goods: high demand on content causes a high demand on access bandwidth and low bandwidth availability causes a low demand on content. In spite of the two services being complementary, the ability of the ISPs to monetize the user surplus has been significantly lower than that of the CPs. It can be argued that this is due to a combination of market forces (average revenue per user has been stubbornly constant, and low, for many years in most markets) and regulations (e.g., net neutrality and others akin to common carrier laws). 

The asymmetric economic system against the ISPs outlined above is compounded by the march of technology and resource demand.  While the content is becoming more demanding on the network resources, the pressure to provide that last mile capacity is on the ISPs. Thus the ISPs have to continuously invest in the access infrastructure to both upgrade to more modern technologies and to expand capacity. Since the CPs appear to be getting a larger share of the consumer surplus, there is an argument, at least from some of the ISPs, that the CPs need to contribute to investments in the access infrastructure that they so significantly profit from. If indeed the CPs have to pay for the public ISP infrastructure, the natural question is to understand the incentive structures that should be created, even as the market forces and the regulatory structure are maintained. 

\subsection{Related Literature}
Early work on addressing the issue of CPs paying for the access centred around allowing ISPs to be non neutral and provide differentiated service quality to the CPs, e.g.,  \cite{Choi10,Economides12a,Economides12b}. Another strain of work considered non neutrality of ISPs where they provide differentiated pricing via schemes like sponsored data and zero rating, e.g., \cite{Schewick14,Rossini15,Phalak19,Vyavahare22}. With net neutrality laws becoming either \textit{de jure} or \textit{de facto} in many markets, these do not provide us any insights on how to make the CPs pay for access infrastructure. A widely used practice that has emerged is for CPs, including content distribution networks (CDNs), to peer with ISPs over a direct link and cache their content deep inside the ISP. Such peering interconnections and caching contribute to reducing cost for ISPs while effectively favouring the traffic from the peering ISP over those that do not peer.  This has been explored in, e.g., \cite{Patchala21}. Another scheme proposed to contribute to ISP revenues is a profit sharing model between the ISP and the CPs \cite{Bouvard22}. In  \cite{Kalvit2019}, the voluntary payment by the CPs to the neutral ISP is treated like a game to provide common good (see, e.g., \cite{Varian10} for a definition) and the total contribution is analysed under different types of games. This paper explores that idea further by including a private investment, in addition to the public investment like in \cite{Kalvit2019}, that the CPs can make to improve the experience of their users. We develop this idea further in the following subsection. 

\subsection{Problem Setup and Preview}
Consider a system with a neutral ISP and several CPs that together serve a fixed user population. The consumption of the contents of a CP depends on the quality of service (QoS) that users experience which in turn depends on the quality of the access network provided by the ISP. The CPs can make a voluntary contribution to the ISP for it to invest in improving the access network infrastructure; an improved access network will contribute to increased content consumption and hence increased revenues to CPs. This voluntary contribution will be called the public investment of the CP. Since the ISP is neutral, all the CPs can benefit from this public investment of some of the CPs and thus there is the possibility of a tragedy of commons in which case none of the CPs will make public investment. 

In addition to collectively investing in the public infrastructure, each CP can also individually invest in its private infrastructure. This private investment could include, better caching, encoding techniques, better connectivity of their servers to the user population, etc. This investment will improve the QoS to the users of only this CP and provide a consequent increase in its revenues. Recently, many CPs have been making public and private investments. For example, Google has rolled out free Wi-Fi stations in many public places within India. Further, it has also invested in building its content delivery network that provides better caching capabilities.

The premise for the model is similar to that in \cite{Kalvit2019} but we add the significantly more realistic framework of the CPs also investing in their own systems. The two-part investment decision adds significantly more complexity to the analysis and also provides for more interesting analyses. While there are several obvious questions in this setup, we will be interested in analysing the resulting public investment under various strategic interactions between the CPs. 
 
In the next section, we describe the system model and develop some preliminary results that will be used in later sections. In Section~\ref{sec:Centralized}, we consider a centralised decision to choose the private and public investments of each of the CPs to maximise the sum utilities of the CPs as collective. In Section~\ref{sec:Nash}, we analyse the system when the CPs decide their investments as the outcome of a non cooperative game and characterise the public and private investments at Nash equilibrium. In Section~\ref{sec:Numericals}, we present some numerical results and discuss the results. Some proofs are carried in the appendix.

\section{System Model}\label{sec:Model}
There are $N$ CPs and one ISP that together provide Internet services to a fixed user population. There is a basic infrastructure that is present at the ISP and the CPs and the interest here is in new investments that need to be made to improve the quality of service for the users. Each CP can make a \emph{public investment} towards the improvement of the infrastructure of the neutral ISP; users of all the CPs see an improvement in the service quality from this investment. A CP can also invest in its private infrastructure to improve the service quality to only its own users and this is the \emph{private investment.} Let $q_n$ be the public investment and $p_n$ be the private investment of CP $n,$ for all $n \in \mathcal{N}$, where $\mathcal{N}:=\{1, 2, \ldots, N\}$. Let $Q := \sum_{n=1}^N q_{n}$ be the total public investment and the $P := \sum_{n=1}^N p_n$ be the total private investment. Our objective in this paper is to evaluate the relation between $P$ and $Q$ under different assumptions of interaction between the CPs. Specifically, we will be interested in, among others, the  measure
\begin{equation}
    \gamma := \frac{Q}{P} \label{eq:gamma}
\end{equation}
that we call the public-private trade-off. 

The public and private investments lead to an improvement in the overall QoS experience of the users, a consequent increase in traffic, and hence, a revenue gain to the CPs. Further, since the ISP operates in the net neutrality regime, the gain is a function of the total public investment. A total public investment of $Q$ and a private investment of $p_n$ results in an increased consumption of the content of CP $n$ by an amount $g_n(Q+h_{n}(p_{n})),$ where both $g_n(\cdot)$ and $h_n(\cdot)$ are concave increasing functions. This is a reasonable assumption because the marginal benefit from increasing investments will decrease. 
Further, $h_{n}(\cdot)$ is such that the inequality 
\begin{equation}
    h_{n}(p_{n}) \geq p_{n} \label{eq:hn_cond}
\end{equation}
holds for some $p_{n}$ implying 
$$g_{n}(Q_{-n} + q_{n} + h(p_{n})) \geq g_{n}(Q_{-n} + q_{n} + p_{n})$$
for some $p_{n}$ and $q_{n}.$ Here $Q_{-n} := \sum_{m=1, m \neq n}^{N} q_{m}.$
This requirement only means that there exists some $(p_n, q_n)$ for which a CP will make non-zero public and private investments, a reasonable technical assumption. 
 
\begin{remark}
The function $g_{n}(\cdot)$ characterizes the amount of increased content consumption of CP $n$ due to the total public investment and its private investment.
A possible view of $h_n(\cdot)$ is that it is the `equivalent public investment' for a private investment of $p_n.$ 
\end{remark}

Denoting the revenue per unit traffic for CP $n$ to be $r_n,$ the surplus, or the utility function for CP $n,$ denoted by $U_{n},$ is  
\begin{equation}
    U_{n}(q_{n}, p_{n}) = r_{n}g_n(Q+h_n(p_{n})) - (p_{n} + q_{n}) , \forall n \in \mathcal{N} . \label{eq:Un1}
\end{equation}

\begin{remark}
    Since $g_n(\cdot)$ and $h_n(\cdot)$ are concave increasing, it is easy to show that $U_n$ is concave in both $p_n$ and $q_n.$ This is not surprising  because the marginal revenue gains from either the public or the private investment should eventually plateau and the net gain will become negative. 
\end{remark}  

In the following sections, we will determine the $p_n$ and $q_n,$ $n \in \mathcal{N}$, when the CPs make their choices under two interaction models---a centralized allocation and a non-cooperative game. In the process, we also obtain the $\gamma$ and the utility functions of the different players under these interactions. In the rest of the paper, we use the following forms of $g_n(\cdot)$ and $h_n(\cdot)$ for analytical tractability.
\begin{equation}
\label{assm:g_h} 
g_n(x) = a_n \log(1+x) \text{ and } h_n(x)= b_n\sqrt{x}\, \,\,\,\,\, \forall \ \ x \geq 0 .
\end{equation}
Further, we impose that $b_{n} \geq 1$ to meet the structural requirement discussed in (\ref{eq:hn_cond}).
Substituting (\ref{assm:g_h}) into (\ref{eq:Un1}), we get
\begin{equation}
    U_{n}(q_{n}, p_{n}) = r_{n}a_{n} \log(1+Q+b_n \sqrt{p_{n}}) - (p_{n}+q_{n}) \, . 
    \label{eq:Un}
\end{equation}

\section{Centralized Allocation} 
\label{sec:Centralized}
First, let us consider a centralised allocation where both the $Q$ and the $p_n$ are determined centrally to maximize the sum utilities of all the CPs  
\begin{equation}
    U_{C}(Q, \{p_n\}) := \sum_{n=1}^{N} \left( r_{n}a_{n} \log(1+Q+b_{n}\sqrt{p_{n}}) - p_{n} \right) - Q.   \label{eq:utility}
\end{equation}
This is an idealisation and will be used to get the best case scenario that will in turn be used as a standard to compare against. The optimization problem in this \emph{centralized} allocation will be 
\begin{align}
    &\max_{Q, \{p_{n}\}} \ \ \sum_{n=1}^{N} (r_{n}a_{n}\log(1+Q+b_{n}\sqrt{p_{n}}) - p_{n}) - Q , \nonumber \\
   \text{s. t. } & p_{n} \geq 0 ~\forall n \in \{1, \ldots, N\} , \nonumber \\
   & Q \geq 0 . \label{eq:centralized_problem}
\end{align}
Towards solving \eqref{eq:centralized_problem}, we first present the following lemma that characterizes the optimum $p_n$ for a given $Q.$ 
\begin{lemma}\label{lem:opti_pn}
    The optimum value of $p_{n}$ for a given value of $Q$, denoted by $p_{n}^{\ast}(Q)$, is given by
    \begin{equation}
        p_{n}^{\ast}(Q) = \left(\frac{\sqrt{(1+Q)^{2}+2b_{n}^{2}r_{n}a_{n}}-(1+Q)}{2b_{n}}\right)^{2} . \label{eq:opti_pn}
    \end{equation}
\end{lemma}

\begin{remark}
Lemma \ref{lem:opti_pn} says that $p_{n}^{\ast}(Q)$ is strictly positive for all $Q \geq 0$ and is a convex decreasing function of $Q$.
\end{remark}
Thus, given a fixed total public investment, a CP can always increase its utility by making private investment, however small. Further, because of the concavity of the utility function, the marginal investment that can improve its utility decreases with increasing $Q.$   

The optimum total public investment with \emph{centralized allocation}, i.e., the solution to \eqref{eq:centralized_problem}, denoted by $Q_{C}^{\ast}$, is characterised by the following lemma. 

\begin{lemma}\label{lem:lemma_coop}
$Q_{C}^{\ast}$ is obtained by solving 
\begin{equation}
    \sum_{n=1}^{N} \frac{\sqrt{(1+Q)^{2}+2b_{n}^{2}r_{n}a_{n}}}{b_{n}^{2}} = (1+Q)\left(\sum_{n=1}^{N} \frac{1}{b_{n}^{2}}\right) + 1. \label{eq:lemma_coop}
\end{equation}
\end{lemma}
Lemma \ref{lem:lemma_coop} implies 
\begin{equation}
    Q_{C}^{\ast}>0 \iff \sum_{n=1}^{N}\frac{\sqrt{1+2b_{n}^{2}r_{n}a_{n}}}{b_{n}^{2}} > 1 + \sum_{n=1}^{N} \frac{1}{b_{n}^{2}}. \label{eq:non-zero_cond}
\end{equation}

Denoting the public-private trade-off under \emph{centralized allocation} by $\gamma_{C},$ we have the following theorem whose proof is provided in the appendix. 
\begin{theorem} \label{thm:coop_gamma}
    For {centralized} allocation of $\{p_n\}$ and $Q,$  
    \begin{equation}
        \gamma_{C} = \frac{2Q_{C}^{\ast}}{\sum_{n=1}^{N}r_{n}a_{n} - (1+Q_{C}^{\ast})} . \label{eq:lemma_coop_gamma}
    \end{equation}
\end{theorem}

\section{Non-Cooperative Interaction}
\label{sec:Nash}
We now consider the case when the CPs choose the $p_n$ and $q_n$ strategically where they do not coordinate and choose these quantities to selfishly maximize their utility. We begin the analysis by first using Lemma \ref{lem:opti_pn} and rewriting $U_{n}$ as 
\begin{eqnarray}
    U_{n} & = & r_{n}a_{n}f_{n}(Q) - p_{n}(Q) - q_{n} 
    \label{eq:Un_2} \\
\mbox{where }  f_{n}(Q) & = & \log\left(\frac{\sqrt{(1+Q)^{2}\!+\!2b_{n}^{2}r_{n}a_{n}}\!+\!(1\!+\!Q)}{2}\right) .
\nonumber 
\end{eqnarray}

Since $U_{n}$ is a function of $Q=\sum_n q_n, $ and the latter are the actions of the CPs, we determine the {Nash equilibrium} choice of $\{q_n\}$ as follows. Let $q_{n}^{\ast}$ denote the equilibrium public investment maximizing the utility of CP $n$. Then, a {pure strategy Nash equilibrium} exists if and only if 
\begin{equation}
    U_{n}(q_{n}^{\ast}, Q_{-n}^{\ast}) \geq U_{n}(q_{n},Q_{-n}^{\ast})
\end{equation}
holds for $\{q_{n}\}_{n \in \mathcal{N}}$ where $Q_{-n}^{\ast}=\sum_{m \in \mathcal{N}, m \neq n} q_{m}^{\ast}.$ In other words, $\{q_{n}^{\ast}\}_{n \in \mathcal{N}}$ is a {Nash equilibrium} public investment tuple that ensures
\begin{equation}
    q_{n}^{\ast} = \argmax_{q_{n}} \, U_{n}(q_{n},Q_{-n}^{\ast})~\forall n \in \mathcal{N}. 
\end{equation}
Now, using $Q_{N}^{\ast} := \sum_{n \in \mathcal{N}} q_{n}^{\ast},$ we  present the following theorem whose proof is in the appendix.  
\begin{theorem} \label{thm:NC}
    For {non-cooperative} interaction, the following holds.
    \begin{itemize}
        \item If $\max \limits_{n \in \mathcal{N}} \left(r_{n}a_{n} - \frac{b_{n}^{2}}{2}\right) \leq 1$, $q_{n}^{\ast}=0$ for all $n \in \mathcal{N}$, and hence, $Q_{N}^{\ast}=0$.
        \item If $\max \limits_{n \in \mathcal{N}} \left(r_{n}a_{n} - \frac{b_{n}^{2}}{2}\right)>1$, let 
        \begin{equation}
            \mathcal{M} := \left\{m \in \mathcal{N}: m = \argmax\limits_{n \in \mathcal{N}}\left(r_{n}a_{n}-\frac{b_{n}^{2}}{2}\right)\right \}. \label{eq:set_M}
        \end{equation}
        Then, $q_{n}^{\ast} \geq 0$ for all $n \in \mathcal{M}$ and $q_{n}^{\ast}=0$ for all $n \in \mathcal{N} \setminus \mathcal{M}$. Further, $1+Q_{N}^{\ast} = \max \limits_{n \in \mathcal{N}} \left(r_{n}a_{n} - \frac{b_{n}^{2}}{2}\right)$.
    \end{itemize}
\end{theorem}

An important conclusion from Theorem \ref{thm:NC} is that only the CP which has the maximum value of $r_{n}a_{n}-\frac{b_{n}^{2}}{2}$ contributes to the public investment and the remaining CPs free-ride on this. If more than one CP has the maximum value of $r_{n}a_{n}-\frac{b_{n}^{2}}{2}$, i.e., some symmetry exists between the CPs, $q_n >0$ for all such CPs. All the CPs though invest in their private infrastructure and the optimum value of $p_{n}$ can be obtained by substituting $Q$ by $Q_{N}^{\ast}$ in \eqref{eq:opti_pn}. 

Let $\eta$ denote the price of anarchy defined as
\begin{equation}
    \eta := \frac{Q_{C}^{\ast}}{Q_{N}^{\ast}} . \label{eq:eta}
\end{equation}
Further, let $\gamma_{N}$ denote the public-private trade-off in the case of the {non-cooperative} interaction. The following theorem provides insights on $\eta$ and $\gamma_{N}.$
\begin{theorem}\label{thm:PoA}
    For \emph{Non-Cooperative} interaction, the following holds.
    \begin{itemize}
        \item If $\max\limits_{n \in \mathcal{N}} \left(r_{n}a_{n}-\frac{b_{n}^{2}}{2}\right) \leq 1$, $\eta$ is unbounded and $\gamma_{N}=0$.
        \item If $\max\limits_{n \in \mathcal{N}} \left(r_{n}a_{n} - \frac{b_{n}^{2}}{2}\right) > 1$, $\eta > 1$. For the special case of $|\mathcal{M}|=|\mathcal{N}|$, 
        \begin{equation}
            \gamma_{N} = \frac{Q_{N}^{\ast}}{\sum \limits_{n \in \mathcal{N}} \frac{b_{n}^{2}}{4}} . \label{eq:gamma_n}
        \end{equation}
    \end{itemize}
\end{theorem}
From Theorem~\ref{thm:PoA} it is clear that $\eta > 1.$ 

Let $U_{N}$ denote the sum utility of all the CPs in the {non-cooperative interaction}. Akin to the derivation of \eqref{eq:utility}, we obtain $U_{N}$ as
\begin{equation}
    U_{N} = \sum_{n \in \mathcal{N}} \left(r_{n}a_{n}\log(1+Q_{N}+b_{n}\sqrt{p_{n}}) - p_{n}\right) - Q_{N} . \label{eq:NE_sum_utility}
\end{equation}
Defining $\Gamma$ as 
\begin{equation}
    \Gamma := \frac{U_{C}^{\ast}}{U_{N}^{\ast}},  \label{eq:Gamma}
\end{equation}
we see from Theorem \ref{thm:PoA} that if $\max \limits_{n \in \mathcal{N}} \left(r_{n}a_{n} - \frac{b_{n}^2}{2}\right)>1$, $\Gamma>1$.

\section{Numerical Results}\label{sec:Numericals}
\begin{figure*}
    \centering
    \subfigure[]
    {
        \includegraphics[height=1.6in,width=2.23in]{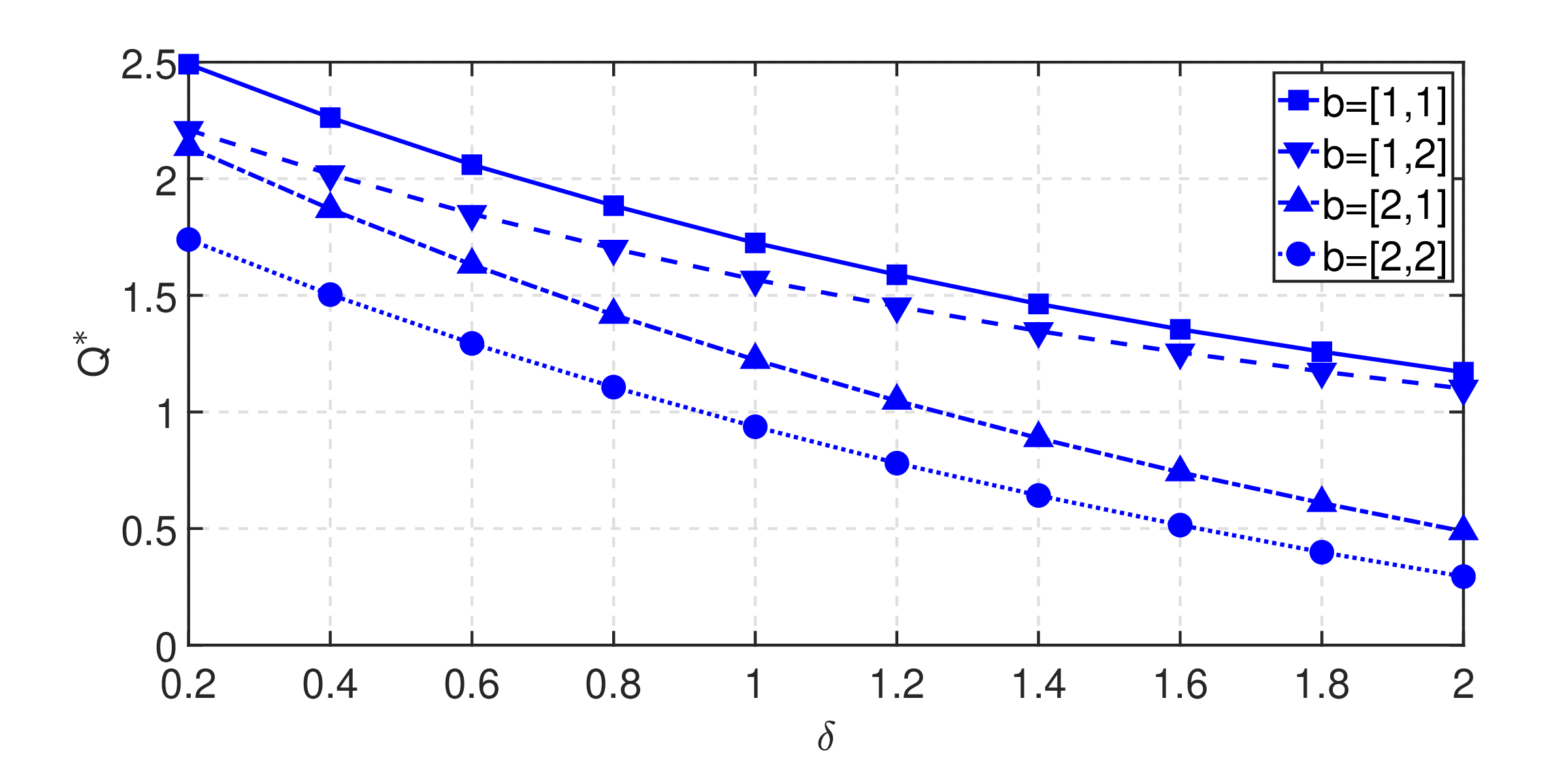}
        \label{fig:cent_pub}
    }
    \subfigure[]
    {
        \includegraphics[height=1.6in,width=2.23in]{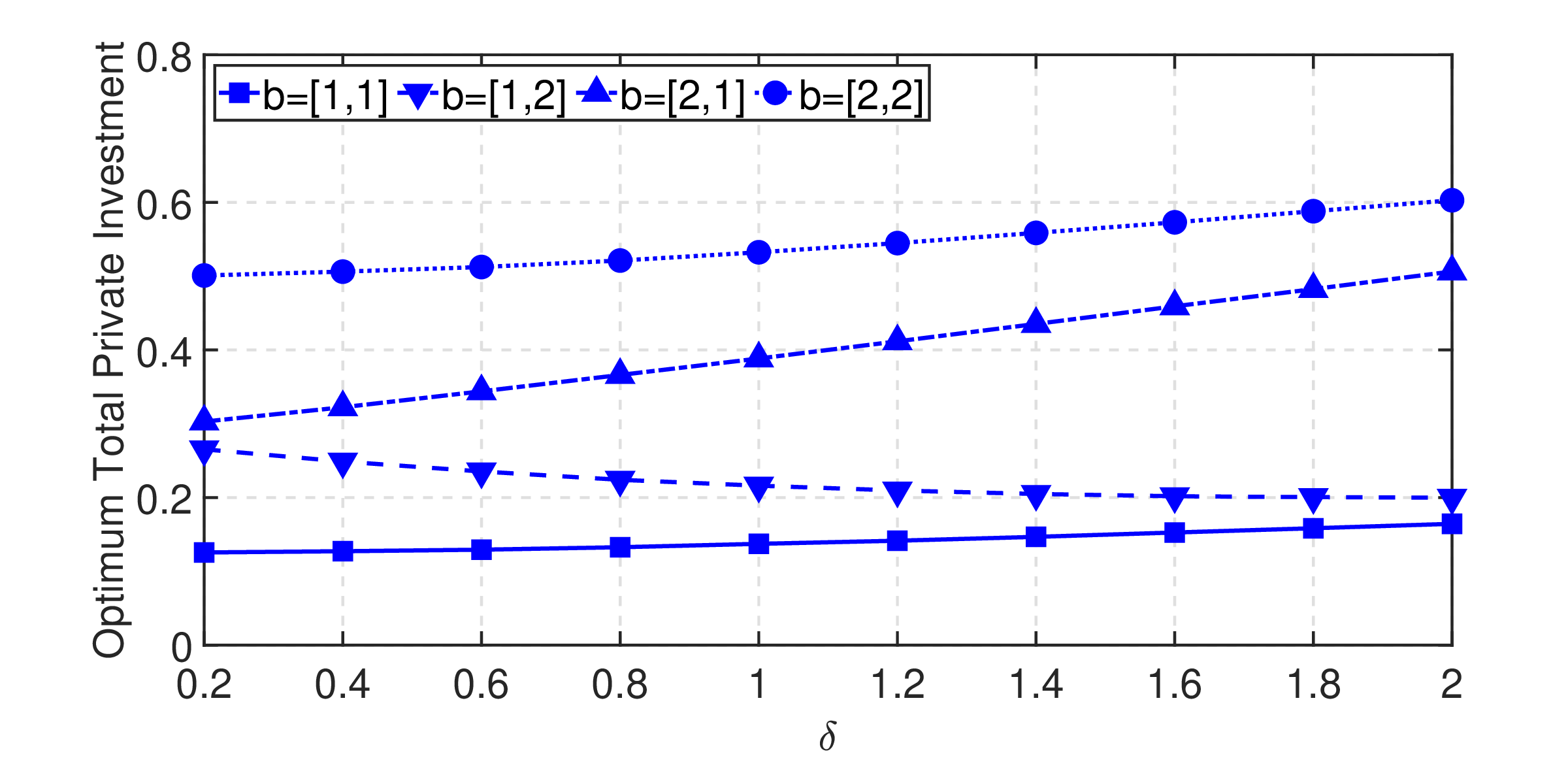}
        \label{fig:cent_priv}
    }
    \subfigure[]
    {
        \includegraphics[height=1.6in,width=2.23in]{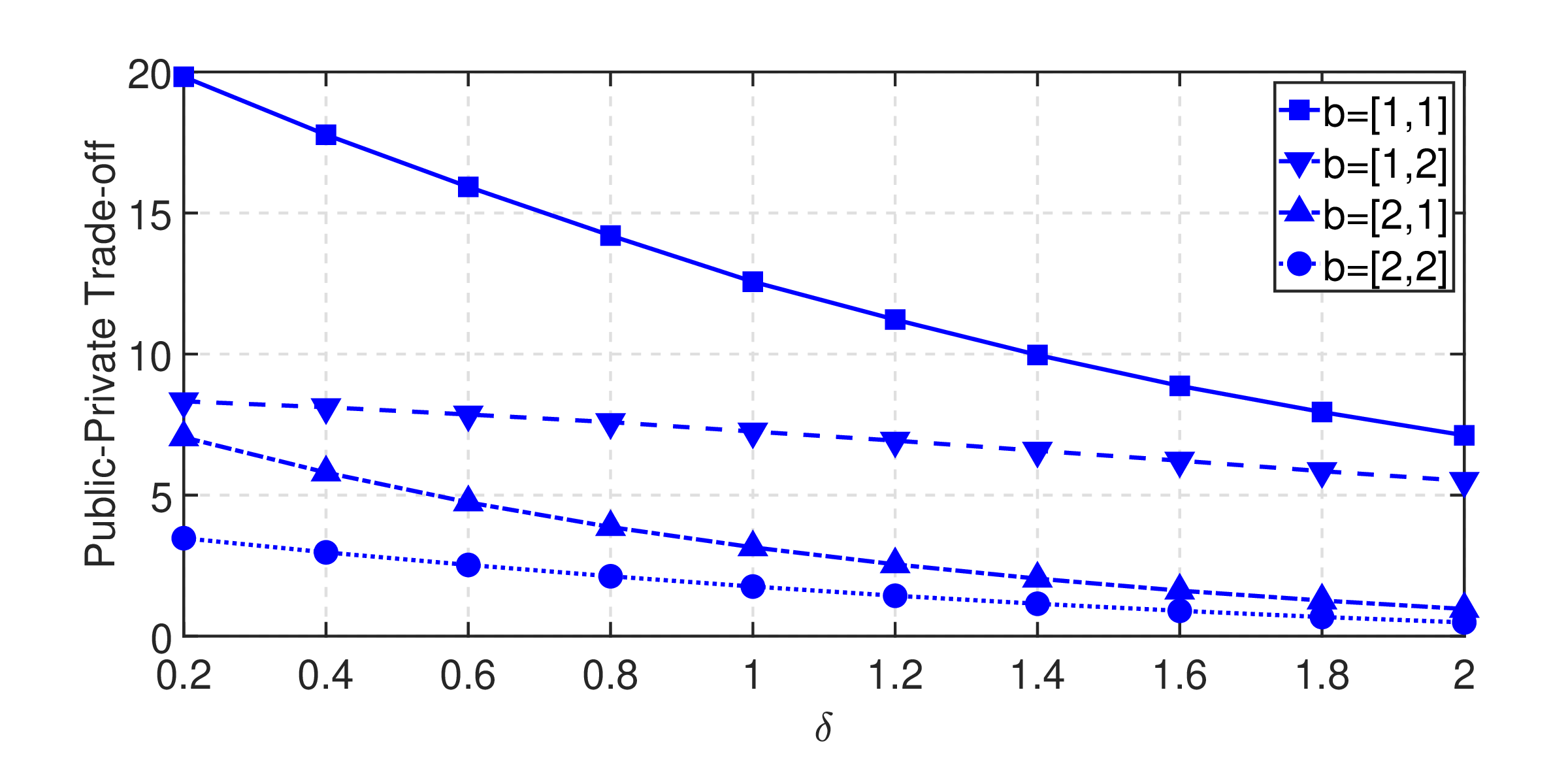}
        \label{fig:cent_ppt}
    }
    \caption{Variation of optimum total public investment, optimum total private investment, and public-private trade-off against different values of $\delta$ and $\textbf{b}$ for \emph{centralized} game and $N=2$ in (a), (b), and (c), respectively.
    }
    \label{fig:centralized}
\end{figure*}
\begin{figure*}
    \centering
    \subfigure[]
    {
        \includegraphics[height=1.6in,width=2.23in]{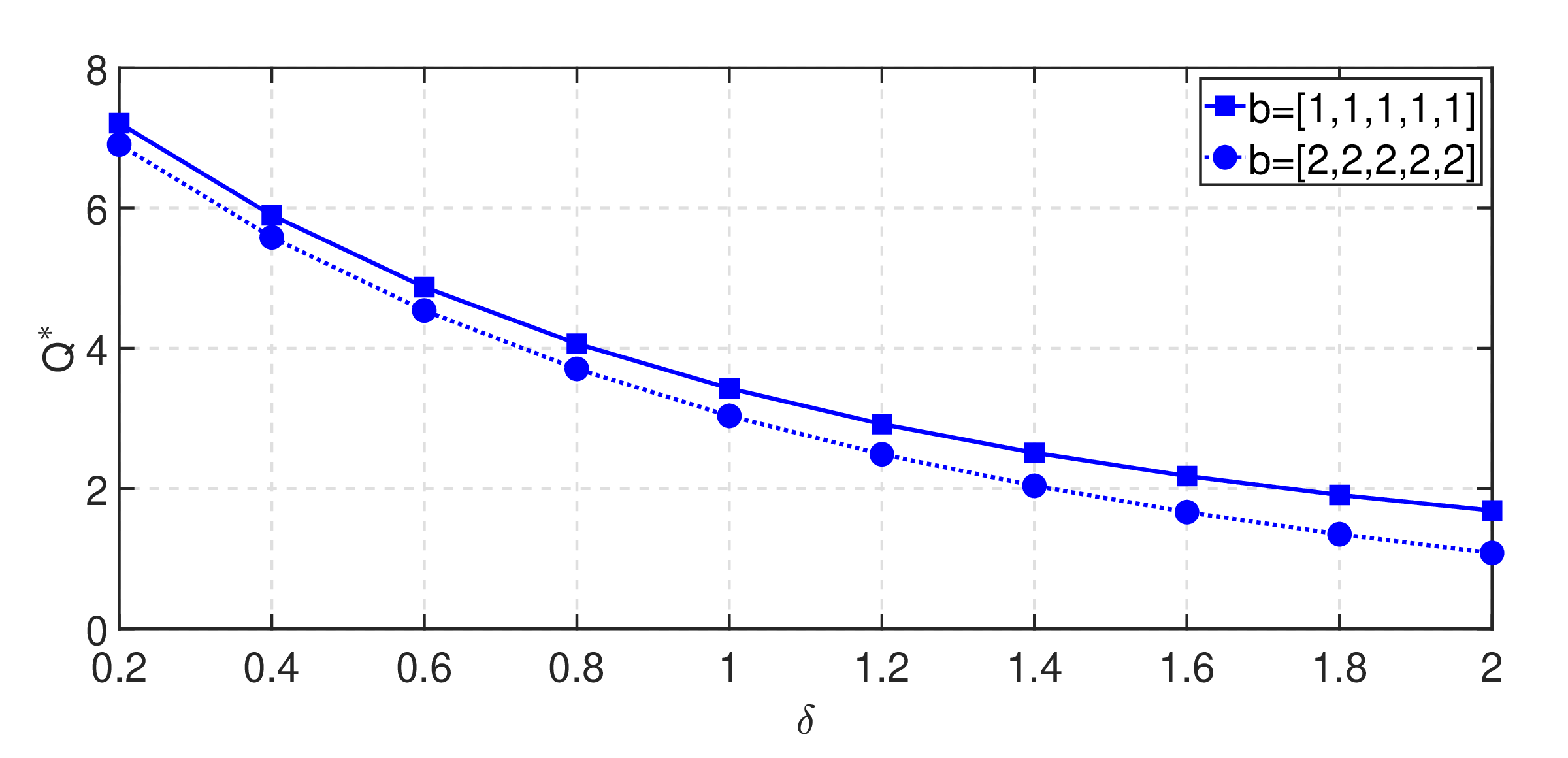}
        \label{fig:cent_pub_N5}
    }
    \subfigure[]
    {
        \includegraphics[height=1.6in,width=2.23in]{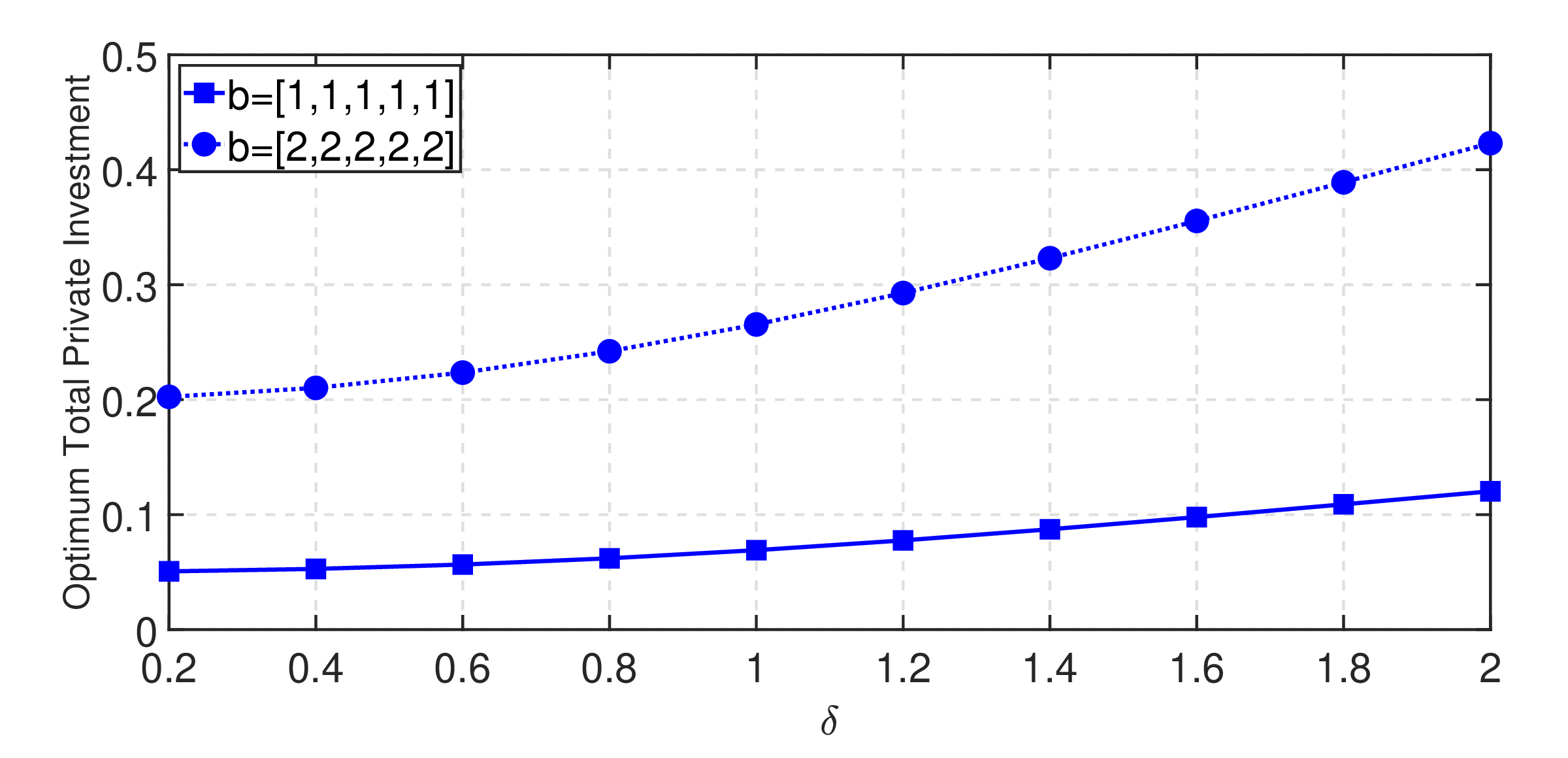}
        \label{fig:cent_priv_N5}
    }
    \subfigure[]
    {
        \includegraphics[height=1.6in,width=2.23in]{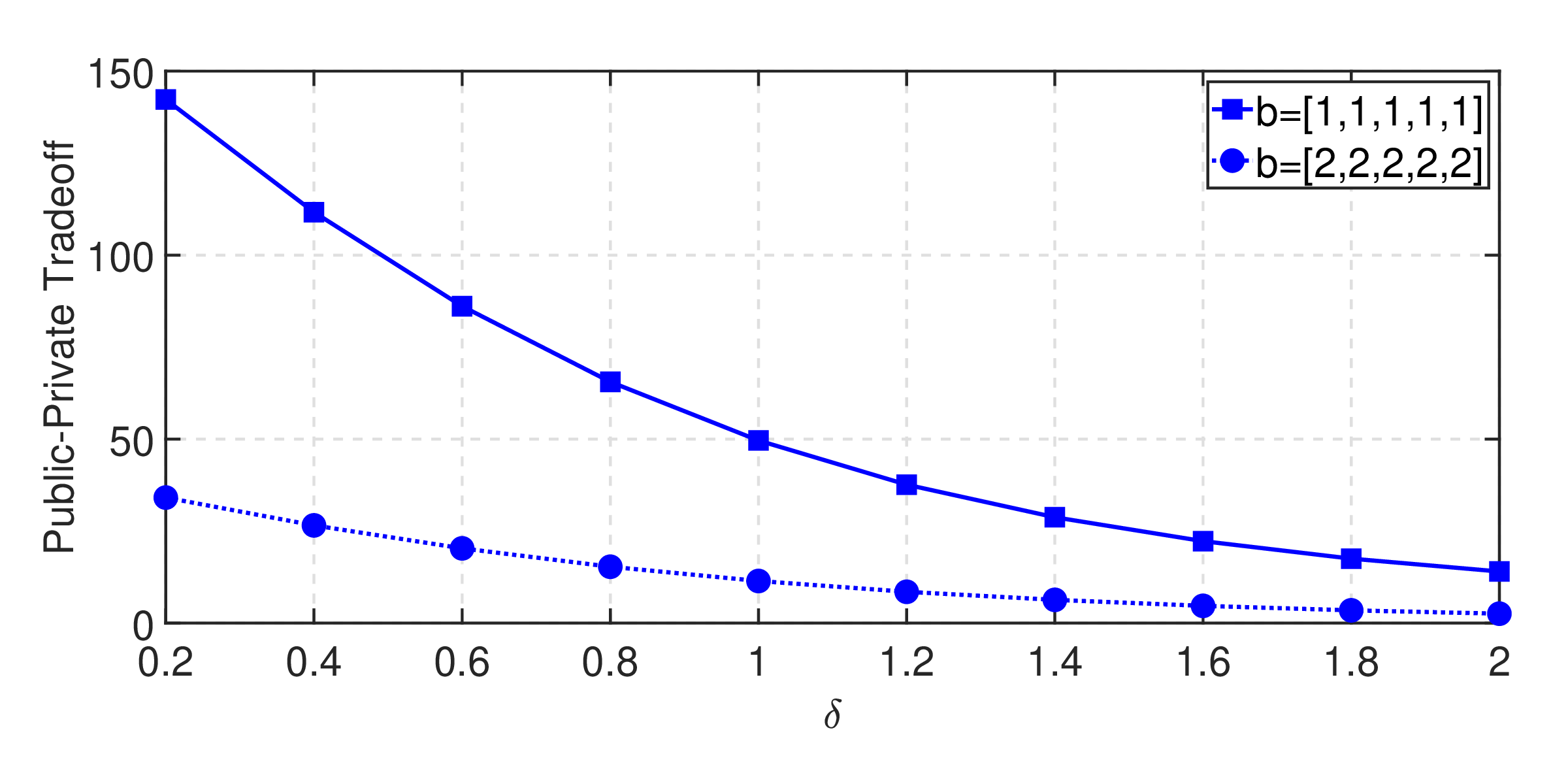}
        \label{fig:cent_ppt_N5}
    }
    \caption{Variation of optimum total public investment, optimum total private investment, and public-private trade-off against different values of $\delta$ and $\textbf{b}$ for \emph{centralized} game and $N=5$ in (a), (b), and (c), respectively.
    }
    \label{fig:centralized_N5}
\end{figure*}
\setlength{\textfloatsep}{0pt}
In this section, we present some numerical results for $N=2$. We consider $r_{n}a_{n} = 2n^{-\delta}$ for all $n=\{1,2\}.$ Here, $\delta$ controls the revenue a CP earns per unit of traffic. Further, the multiplicative factor of $2$ ensures that the optimum total public investment is non-zero for all values of $\delta$, i.e., \eqref{eq:non-zero_cond} holds for all $\delta$. Note that the second CP's revenue per unit of traffic diminishes as $\delta$ increases. The optimization problem formulated in \eqref{eq:centralized_problem} is solved using Pyomo.

Fig. \ref{fig:centralized} presents the variation of the optimum total public investment, optimum total private investment, and public-private trade-off against different values of $\delta$ and $\textbf{b}$ for {centralized} game in Fig. \ref{fig:cent_pub}, \ref{fig:cent_priv}, and \ref{fig:cent_ppt}, respectively. We observe from Fig. \ref{fig:cent_pub} that the optimum total public investment, i.e., $Q^{\ast}$, decreases as $\delta$ increases. This follows from the fact that the revenue of CP 2 per unit of traffic is a decreasing function of $\delta$. Since the revenue of CP 1 per unit of traffic does not change with $\delta$, the decrease in $Q^{\ast}$ leads to an increase in the optimum private investment of CP 1. However, since the revenue of CP 2 per unit of traffic decreases with $\delta$, the optimum private investment of CP 2 decreases with $\delta$. This explains the mixed trend of the optimum total private investment, which is the sum of the optimal private investments of both CPs, against different values of $\delta$ as observed in Fig. \ref{fig:cent_priv}. In the case of $\textbf{b}=[2, 2]$, both the CPs are incentivized to invest more in private than public. Therefore, we observe that the optimal total public and private investment are lowest and highest for $\textbf{b}=[2, 2]$ in Fig. \ref{fig:cent_pub} and \ref{fig:cent_priv}, respectively. A similar explanation holds for the highest and lowest values of optimum total public and private investments for the case of $\textbf{b}=[1, 1]$ in Fig. \ref{fig:cent_pub} and \ref{fig:cent_priv}, respectively. 
This also translates into the highest public-private trade-off for $\textbf{b}=[1, 1]$ in Fig. \ref{fig:cent_ppt}.
Since in $\textbf{b}=[1, 2]$, only CP 1 prefers to invest more in public than private, we observe from Fig. \ref{fig:cent_pub} that the $Q^{\ast}$ is higher for $\textbf{b}=[1, 2]$ than $\textbf{b}=[2, 1]$. The observation of lower optimum total private investment for $\textbf{b}=[1, 2]$ than $\textbf{b}=[2,1]$ in Fig. \ref{fig:cent_priv} also follows from the same explanation. 
For all the cases except $\textbf{b}=[1, 1]$, CPs have invested some share of the public investment in the private infrastructure to maximize their surplus leading to the reduction in the public-private trade-off. 
Fig. \ref{fig:centralized_N5} presents the variation of the optimum total public investment, optimum total private investment, and public-private trade-off against different values of $\delta$ and $\textbf{b}$ for {centralized} game and $N=5$ in Fig. \ref{fig:cent_pub_N5}, \ref{fig:cent_priv_N5}, and \ref{fig:cent_ppt_N5}, respectively. We observe similar trends for the optimum total public and private investments and public-private trade-off for $N=5$ in Fig. \ref{fig:centralized_N5} as that observed for $N=2$ in Fig. \ref{fig:centralized}.
\begin{figure*}
    \centering
    \subfigure[]
    {
        \includegraphics[height=1.5in,width=2.1in]{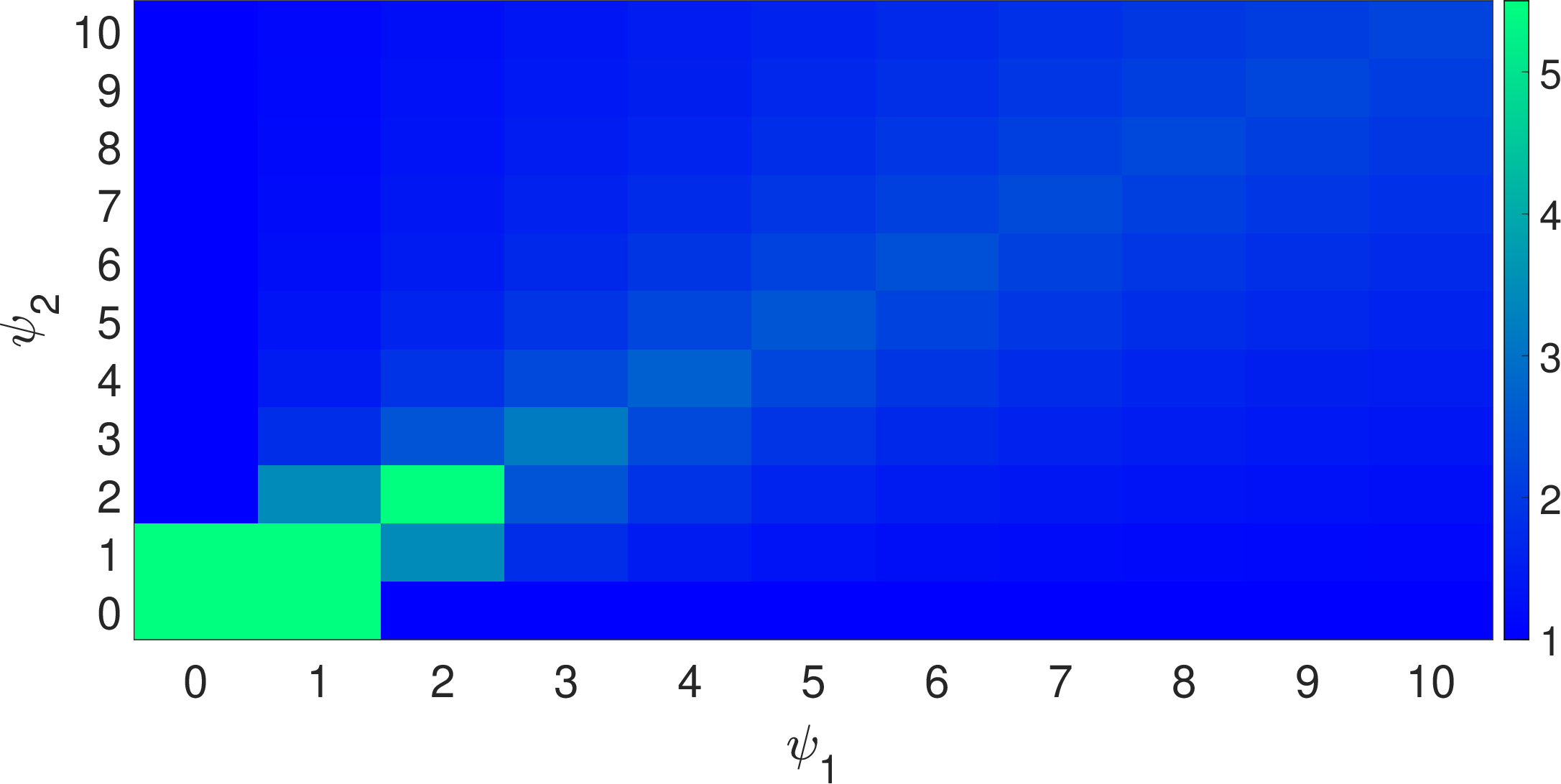}
        \label{fig:poa}
    }
    \subfigure[]
    {
        \includegraphics[height=1.5in,width=2.1in]{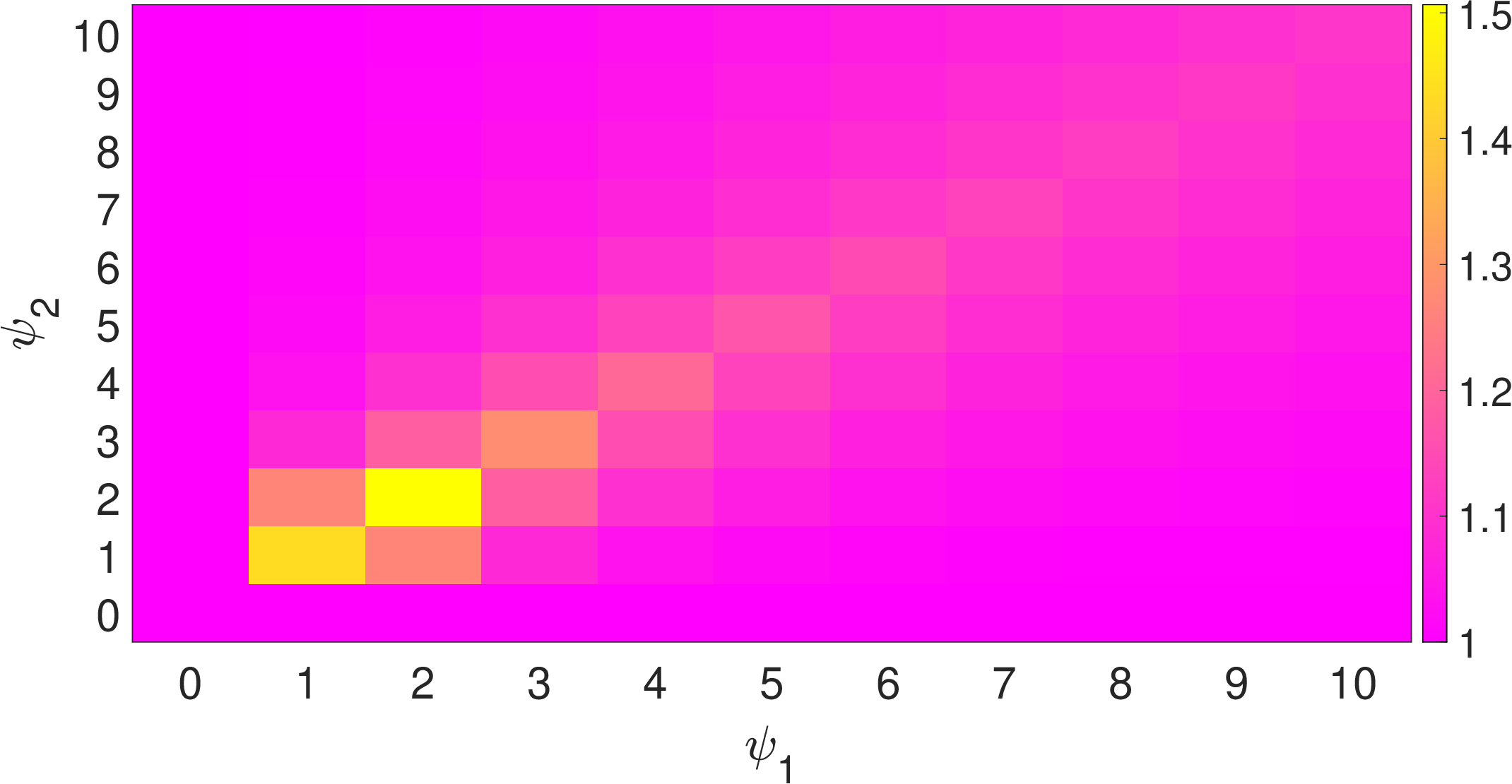}
        \label{fig:gamma}
    }
    \subfigure[]
    {
        \includegraphics[height=1.5in,width=2.1in]{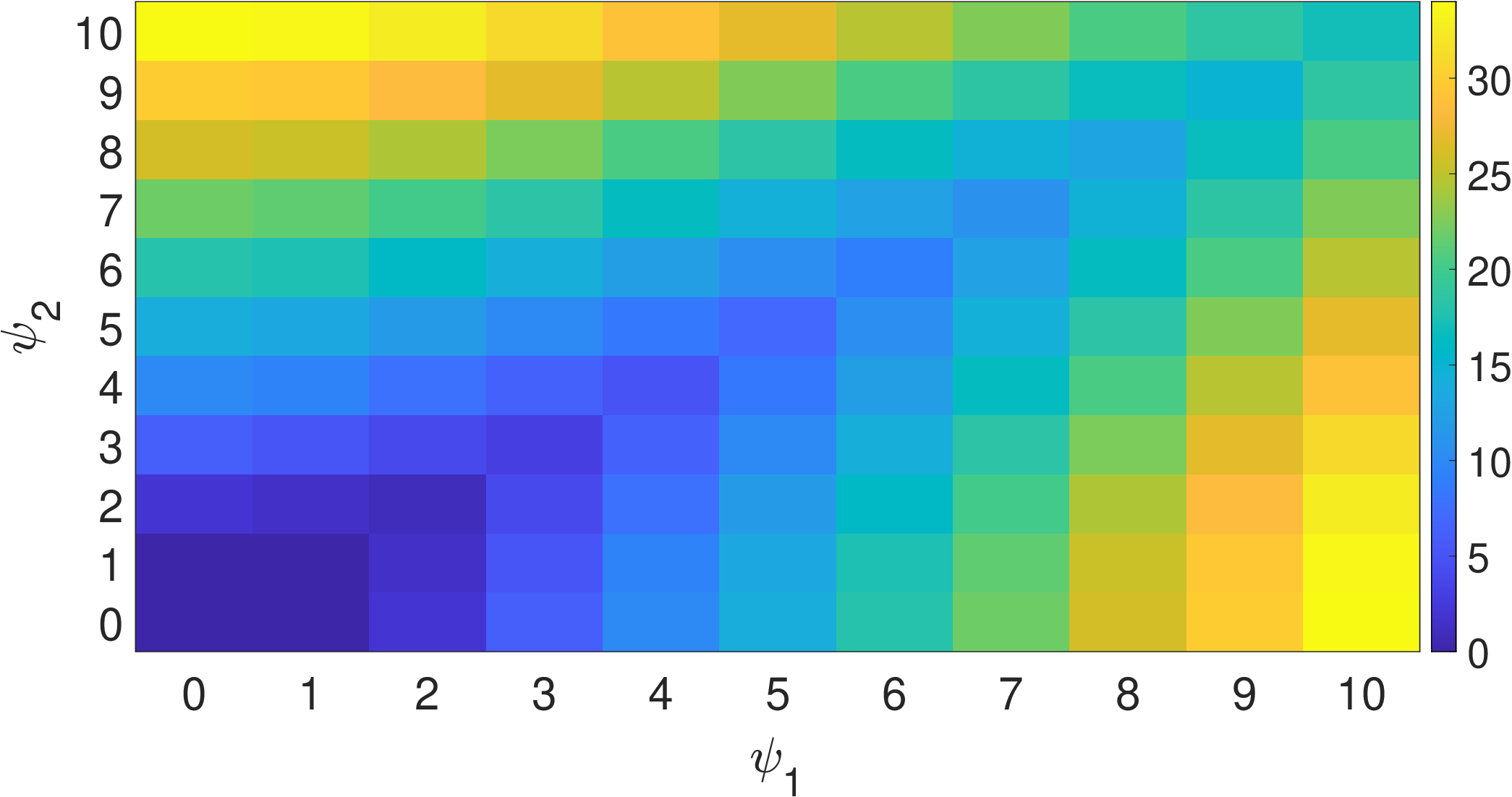}
        \label{fig:NE_ppt}
    }
    \caption{Illustration of the price of anarchy ($\eta$), $\Gamma$, and the public-private trade-off in \emph{non-cooperative interaction} ($\gamma_{N}$) against different values of $\psi_{n}(=r_{n}a_{n})$ for $N=2$ and $\textbf{b}=[1, 1]$ in (a), (b), and (c), respectively.
    }
    \label{fig:NE}
\end{figure*}
\setlength{\textfloatsep}{0pt}
\begin{figure*}
    \centering
    \subfigure[]
    {
        \includegraphics[height=1.5in,width=2.1in]{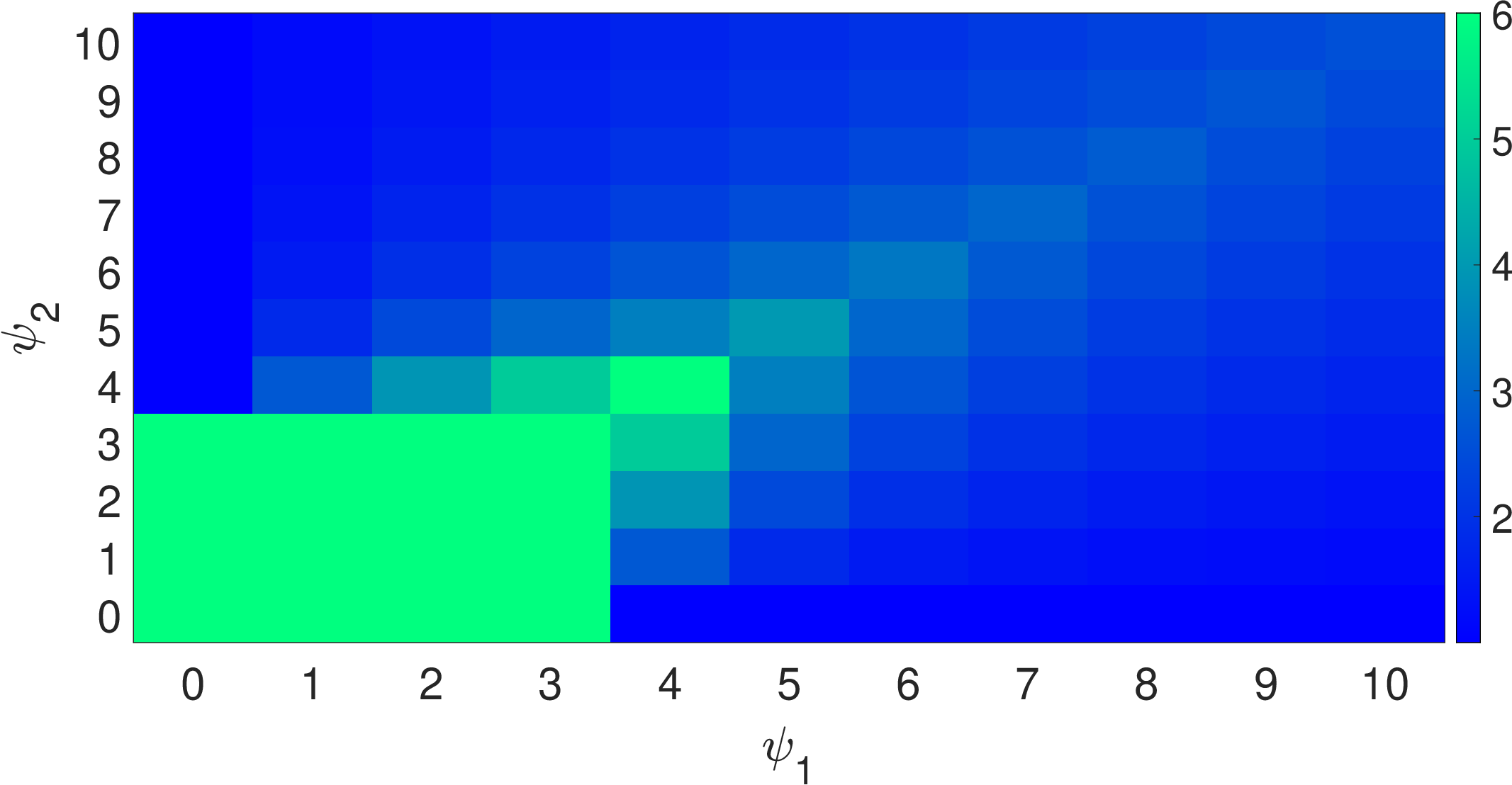}
        \label{fig:poa_b22}
    }
    \subfigure[]
    {
        \includegraphics[height=1.5in,width=2.1in]{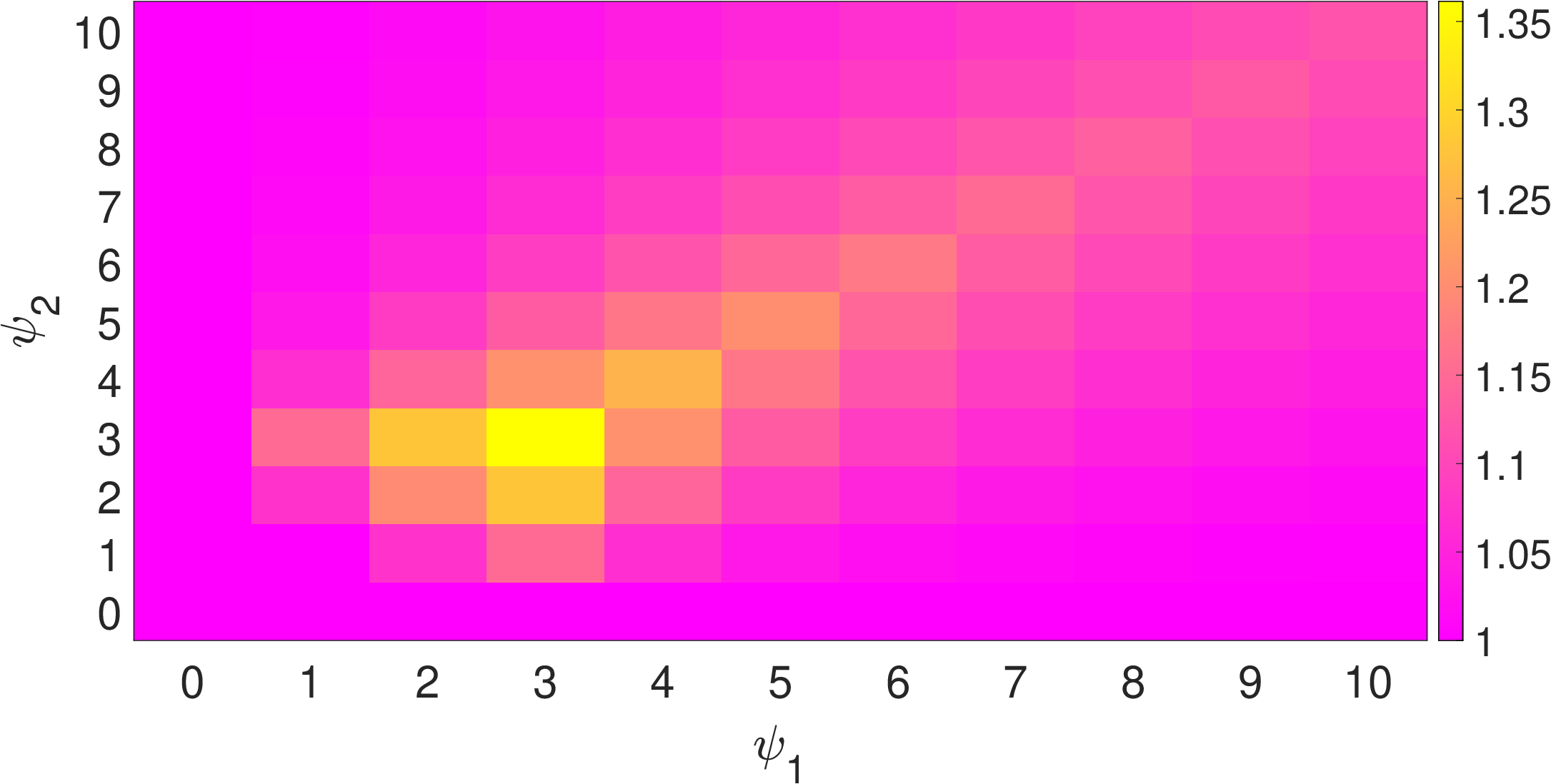}
        \label{fig:gamma_b22}
    }
    \subfigure[]
    {
        \includegraphics[height=1.5in,width=2.1in]{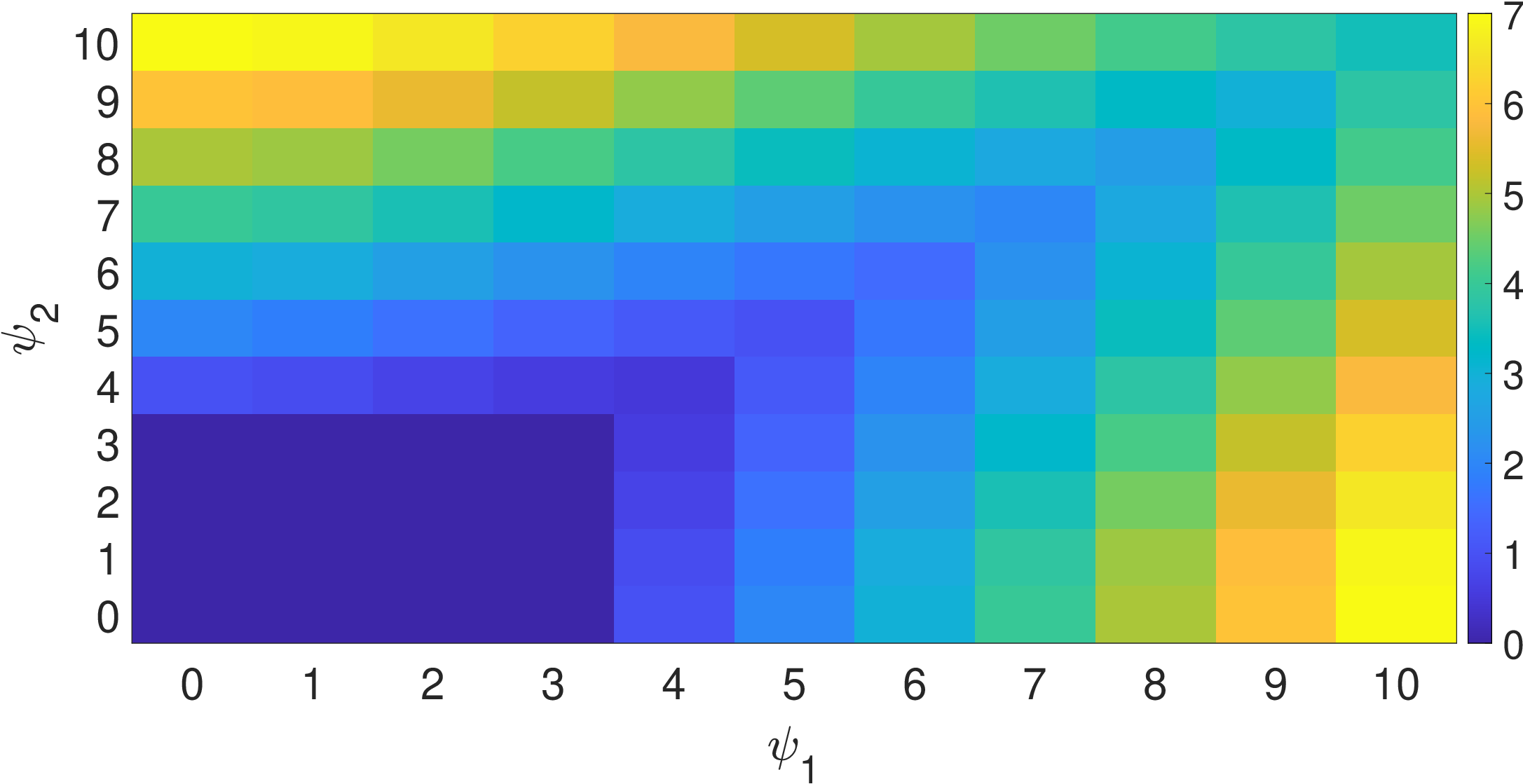}
        \label{fig:NE_ppt_b22}
    }
    \caption{Illustration of the price of anarchy ($\eta$), $\Gamma$, and the public-private trade-off in \emph{non-cooperative interaction} ($\gamma_{N}$) against different values of $\psi_{n}(=r_{n}a_{n})$ for $N=2$ and $\textbf{b}=[2, 2]$ in (a), (b), and (c), respectively.
    }
    \label{fig:NE_b22}
\end{figure*}
\setlength{\textfloatsep}{0pt}

Fig. \ref{fig:NE} illustrates price of anarchy ($\eta$), $\Gamma$, and public-private trade-off for \emph{non-cooperative interaction} ($\gamma_{N}$) against different values of $\psi_{n}(=r_{n}a_{n})$ for $N=2$ and $\textbf{b}=[1, 1]$ in Fig. \ref{fig:poa}, \ref{fig:gamma}, and \ref{fig:NE_ppt}, respectively. 
Similarly, Fig. \ref{fig:NE_b22} illustrates price of anarchy ($\eta$), $\Gamma$, and public-private trade-off for \emph{non-cooperative interaction} ($\gamma_{N}$) against different values of $\psi_{n}(=r_{n}a_{n})$ for $N=2$ and $\textbf{b}=[2, 2]$ in Fig. \ref{fig:poa_b22}, \ref{fig:gamma_b22}, and \ref{fig:NE_ppt_b22}, respectively. 
We observe from Fig. \ref{fig:poa} that $\eta$ is high for the region $(\psi_{1} \leq 1.5, \psi_{2} \leq 1.5)$. This follows from Theorem \ref{thm:PoA} as $\max \limits_{n \in \mathcal{N}} \left(r_{n}a_{n} - \frac{b_{n}^{2}}{2}\right) \leq 1$ in this region. 
Same holds for the region $(\psi_{1} \leq 3, \psi_{2} \leq 3)$ in Fig. \ref{fig:poa_b22} as $\textbf{b}=[2, 2]$. 
In the remaining region, we observe that $\eta>1$ which is consistent with Theorem \ref{thm:PoA}. When $\psi_{1} \approx \psi_{2}$, both CPs invest in public infrastructure leading to a higher value of $Q_{C}^{\ast}$, and hence, $\eta$. However, when there is a large asymmetry between the CPs, only a CP makes a major investment in public infrastructure leading to a lower value of $\eta$.
Similar observations can be made from Fig. \ref{fig:poa_b22}. However, note that the optimum total public investment is less in the case of $\textbf{b}=[2,2]$ than $\textbf{b}=[1,1]$ for both \emph{Centralized Allocation} as well as \emph{Non-Cooperative Interaction}. This is because both the CPs are incentivized to invest more in private than public. Thus, we observe that $\eta$ is slightly higher for $\textbf{b}=[2, 2]$ than $\textbf{b}=[1, 1]$.
In \emph{Centralized Allocation}, the optimum total public investment maximizes the net utility of all the CPs. However, in \emph{Non-Cooperative interaction}, CPs reduce the net public investment by playing selfishly. This results in lower net utility in \emph{Non-Cooperative Interaction} than \emph{Centralized Allocation} as observed in Fig. \ref{fig:gamma}. Similar holds for the case of $\textbf{b}=[2, 2]$ as observed in Fig. \ref{fig:gamma_b22}.
As $\psi_{1}$ increases, $Q_{N}^{\ast}$ increases. Thus, the optimum total private investment decreases, and $\gamma_{N}$ increases as observed in Fig. \ref{fig:NE_ppt}. We observe from Fig. \ref{fig:NE_ppt} and \ref{fig:NE_ppt_b22} that $\gamma_{N}$ significantly reduces for $\textbf{b}=[2, 2]$ than $\textbf{b}=[1, 1]$. This is because both the CPs are incentivized to invest more in private infrastructure than public for $\textbf{b}=[2, 2]$ than $\textbf{b}=[1, 1]$.

\section{Conclusions and Future Works}
With increasing interest by CPs to participate in improving the access network quality, we consider the problem of how CPs can contribute under two kinds of interactions---the idealised centralised allocation and a more realistic strategic model. An interesting finding in the latter case is that at most one CP will contribute to the public infrastructure unless some symmetry exists among the CPs. We expect to investigate this further. We could also consider other models for interactions---cooperative games and a bargaining model. It must be noted that in the bargaining model in \cite{Kalvit2019}, the public investment was higher than in the cooperative setting. It would be interesting to see if this would indeed be the case here. The analysis in this work is limited to one ISP. However, in practice, multiple ISPs provide internet services in a region with different access technologies. So, we plan to look into this tradeoff with multiple ISPs.

\bibliographystyle{IEEEtran}
\bibliography{Globecom_Arxiv} 

\begin{thebibliography}{10}
\providecommand{\url}[1]{#1}
\csname url@samestyle\endcsname
\providecommand{\newblock}{\relax}
\providecommand{\bibinfo}[2]{#2}
\providecommand{\BIBentrySTDinterwordspacing}{\spaceskip=0pt\relax}
\providecommand{\BIBentryALTinterwordstretchfactor}{4}
\providecommand{\BIBentryALTinterwordspacing}{\spaceskip=\fontdimen2\font plus
\BIBentryALTinterwordstretchfactor\fontdimen3\font minus
  \fontdimen4\font\relax}
\providecommand{\BIBforeignlanguage}[2]{{%
\expandafter\ifx\csname l@#1\endcsname\relax
\typeout{** WARNING: IEEEtran.bst: No hyphenation pattern has been}%
\typeout{** loaded for the language `#1'. Using the pattern for}%
\typeout{** the default language instead.}%
\else
\language=\csname l@#1\endcsname
\fi
#2}}
\providecommand{\BIBdecl}{\relax}
\BIBdecl

\bibitem{Choi10}
J.~P. Choi and B.~C. Kim, ``Net neutrality and investment incentives,''
  \emph{{RAND} Journal of Economics}, vol.~41, no.~3, pp. 446--471, Autumn
  2010.

\bibitem{Economides12a}
N.~Economides and J.~Tag, ``Network neutrality on the {I}nternet: A two-sided
  market analysis,'' \emph{Information Economics and Policy}, vol.~24, no.~2,
  pp. 91--104, June 2012.

\bibitem{Economides12b}
N.~Economides and E.~Hermalin, ``The economics of network neutrality,''
  \emph{{RAND} Journal of Economics}, vol.~43, no.~4, pp. 602--629, 2012.

\bibitem{Schewick14}
B.~van Schewick, ``Network neutrality and zero-rating,''
  \url{http://apps.fcc.gov/ecfs/document/view?id=60001031582}, filing at FCC,
  19 February 2014 (Accessed on 10 July 2016).

\bibitem{Rossini15}
C.~Rossini and T.~Moore, ``Exploring zero-rating challenges: Views from five
  countries,'' Public Knowledge Working Paper.

\bibitem{Phalak19}
K.~Phalak, D.~Manjunath, and J.~Nair, ``Zero rating: The power in the middle,''
  \emph{IEEE/ACM Transactions on Networking}, vol.~27, no.~2, pp. 862--874,
  2019.

\bibitem{Vyavahare22}
P.~Vyavahare, J.~Nair, and D.~Manjunath, ``Sponsored data: On the effect of
  {ISP} competition on pricing dynamics and content provider market
  structures,'' \emph{IEEE/ACM Transactions on Networking}, vol.~30, no.~5, pp.
  2018--2031, 2022.

\bibitem{Patchala21}
S.~Patchala, S.~Lee, C.~J. Joo, and D.~Manjunath, ``On the economics of network
  interconnections and its impact on net neutrality,'' \emph{IEEE Transactions
  on Network and Service Management}, vol.~18, no.~4, pp. 4789--4800, 2021.

\bibitem{Bouvard22}
B.~Jullien and M.~Bouvard, ``Fair cost sharing: Big tech vs telcos,'' Toulouse
  School of Economics, Tech. Rep., October 2022.

\bibitem{Kalvit2019}
A.~Kalvit, S.~Pinjani, G.~Kasbekar, D.~Manjunath, and J.~Nair, ``Capacity
  expansion of neutral {ISP}s via content peering charges: {T}he bargaining
  edge,'' \emph{Performance Evaluation}, vol. 133, pp. 43--56, 2019.

\bibitem{Varian10}
H.~Varian, \emph{Intermediate Microeconomics: A Modern Approach}, 8th~ed.\hskip
  1em plus 0.5em minus 0.4em\relax W.~W.~Norton \& Company, 2010.

\end{thebibliography}
\appendix
\subsection{Proof of Lemma \ref{lem:opti_pn}}
\begin{proof}
The partial derivative of $U_C(Q,\{p_{n}\})$, given by (\ref{eq:utility}), with respect to $p_{n}$ is given by
\begin{equation}
    \frac{\partial U_{C}}{\partial p_{n}} = \frac{r_{n}a_{n}b_{n}}{2\sqrt{p_{n}}(1+Q+b_{n}\sqrt{p_{n}})} - 1. \label{eq:partial_pn}
\end{equation}
From (\ref{eq:partial_pn}), we obtain the following quadratic equation in $\sqrt{p_{n}}$ by setting $\frac{\partial U_{C}}{\partial p_{n}}=0$,
\begin{equation}
    b_{n}p_{n} + \sqrt{p_{n}}(1+Q) - \frac{r_{n}a_{n}b_{n}}{2} = 0 . \label{eq:quadratic}
\end{equation}
The roots of (\ref{eq:quadratic}) are given by
\begin{equation}
    \sqrt{p_{n}} = \frac{-(1+Q) \pm \sqrt{(1+Q)^{2}+2b_{n}^{2}r_{n}a_{n}}}{2b_{n}}. \nonumber
\end{equation}
Since $a_{n}$, $r_{n}$, and $b_{n}$ are positive for all $n$, we have $\sqrt{(1+Q)^{2}+2b_{n}^{2}r_{n}a_{n}}>(1+Q)$. Further, $p_{n}$ has to be non-negative for all $n$. This implies the only possible solution of (\ref{eq:quadratic}) is
\begin{equation}
    \sqrt{p_{n}} = \frac{\sqrt{(1+Q)^{2}+2b_{n}^{2}r_{n}a_{n}}-(1+Q)}{2b_{n}} . \label{eq:pn_opti}
\end{equation}
From our discussion above, it follows that $p_{n}>0$ for any value of $Q$. 

To see that the $p_{n}$ is a convex decreasing function of $Q$, we consider $t=\sqrt{p_{n}}$. The partial derivative of $t$ with respect to $Q$ is given by
\begin{equation}
    \frac{\partial t}{\partial Q} = \frac{1}{2b_{n}}\left(\frac{(1+Q)-\sqrt{(1+Q)^{2}+2b_{n}^{2}r_{n}a_{n}}}{\sqrt{(1+Q)^{2}+2b_{n}^{2}r_{n}a_{n}}}\right). \label{eq:partial_t}
\end{equation}
It is easy to see from (\ref{eq:partial_t}) that $t$ is a decreasing function of $Q$. The second order derivative of $t$ with respect to $Q$ is given by
\begin{align}
    \frac{\partial^{2}t}{\partial Q^{2}} &= \frac{1}{2b_{n}\sqrt{(1\!+\!Q)^{2}\!+\!2b_{n}^{2}r_{n}a_{n}}}\left(1 \!-\! \frac{(1\!+\!Q)^{2}}{(1\!+\!Q)^{2}\!+\!2b_{n}^{2}r_{n}a_{n}}\right) \nonumber \\
    &= \frac{r_{n}a_{n}b_{n}}{((1+Q)^{2}+2b_{n}^{2}r_{n}a_{n})^{3/2}} . \label{eq:SOder_t}
\end{align}
From (\ref{eq:SOder_t}), $\frac{\partial^{2}t}{\partial Q^{2}}>0$ which implies that $t$ is a convex function of $Q$.
Since $p_{n}=t^{2}$, we have
\begin{equation}
    \frac{\partial p_{n}}{\partial Q} = 2t\frac{\partial t}{\partial Q} . \label{eq:pn_der}
\end{equation}
Since $t>0$ and $\frac{\partial t}{\partial Q}<0$, it follows that $\frac{\partial p_{n}}{\partial Q}<0$. This implies that the $p_{n}$ is a decreasing function of $Q$. 
The second order derivative of $p_{n}$ with respect to $Q$ is given by
\begin{equation}
    \frac{\partial^{2}p_{n}}{\partial Q^{2}} = 2 \left(\frac{\partial t}{\partial Q}\right)^{2} + 2t\frac{\partial^{2}t}{\partial Q^{2}} . \nonumber 
\end{equation}
Since $t>0$, $\frac{\partial^{2} t}{\partial Q^{2}}>0$, and $\left(\frac{\partial t}{\partial Q}\right)^{2}>0$ for all $Q$, we have $\frac{\partial^{2} p_{n}}{\partial Q^{2}}>0$. Therefore, $p_{n}(Q)$ is a convex decreasing function of $Q$.
This completes the proof.
\end{proof}
\subsection{Proof of Lemma \ref{lem:lemma_coop}}
\begin{proof}
    The partial derivative of $U_{C}$, given in (\ref{eq:utility}), with respect to $Q$ is computed as
    \begin{align}
        \frac{\partial U_{C}}{\partial Q} &= \sum_{n=1}^{N} \frac{r_{n}a_{n}}{1+Q+b_{n}\sqrt{p_{n}}} - 1 , \nonumber
    \end{align}
    Note that $\frac{\partial U_{C}}{\partial p_{n}}$ is given in (\ref{eq:partial_pn}), and hence, using (\ref{eq:pn_opti}), we obtain
    \begin{equation}
        \frac{\partial U_{C}}{\partial Q} = \sum_{n=1}^{N} \frac{2r_{n}a_{n}}{\sqrt{(1+Q)^{2}+2b_{n}^{2}r_{n}a_{n}}+(1+Q)} - 1, \nonumber
    \end{equation}
    which can be simplified to 
    \begin{equation}
        \frac{\partial U_{C}}{\partial Q} = \sum_{n=1}^{N} \left(\frac{\sqrt{(1+Q)^{2}+2b_{n}^{2}r_{n}a_{n}} - (1+Q)}{b_{n}^{2}}\right) - 1 . \label{eq:U_derivative}
    \end{equation}
    Clearly, (\ref{eq:U_derivative}) is the same as (\ref{eq:lemma_coop}).
    If $\left.\frac{\partial U_{C}}{\partial Q}\right\rvert_{Q=0}<0$, it implies $Q^{\ast}_{C}=0$. From (\ref{eq:U_derivative}), we have
    \begin{equation}
       \left. \frac{\partial U_{C}}{\partial Q}\right \rvert_{Q=0} = \sum_{n=1}^{N} \frac{\sqrt{1+2b_{n}^{2}r_{n}a_{n}}-1}{b_{n}^{2}} - 1. \nonumber
    \end{equation}
    Therefore, if $\sum_{n=1}^{N}\frac{\sqrt{1+2b_{n}^{2}r_{n}a_{n}}}{b_{n}^{2}}>1 + \sum_{n=1}^{N} \frac{1}{b_{n}^{2}}$, $\left.\frac{\partial U_{C}}{\partial Q}\right\rvert_{Q=0}>0$ which implies $Q^{\ast}_{C}>0$. Similarly, if $Q^{\ast}_{C}>0$, we have $\left.\frac{\partial U_{C}}{\partial Q}\right\rvert_{Q=0}>0$ which implies $\sum_{n=1}^{N} \frac{\sqrt{1+2b_{n}^{2}r_{n}a_{n}}}{b_{n}^{2}} > 1 + \sum_{n=1}^{N} \frac{1}{b_{n}^{2}}$. This completes the proof.  
\end{proof}
\subsection{Proof of Theorem \ref{thm:coop_gamma}}
\begin{proof}
    From Lemma \ref{lem:lemma_coop}, we know that $Q_{C}^{\ast}$ is obtained by solving (\ref{eq:lemma_coop}). Further, $p_{n}^{\ast}(Q)$ is given by (\ref{eq:opti_pn}). Thus, we have
    \begin{equation}
        p_{n}^{\ast}(Q) = \frac{(1+Q)}{2b_{n}^{2}} + \frac{r_{n}a_{n}}{2} - \frac{(1+Q)\sqrt{(1+Q)^{2}+2b_{n}^{2}r_{n}a_{n}}}{2b_{n}^{2}} \nonumber 
    \end{equation}
    Using (\ref{eq:lemma_coop}), we obtain
    \begin{align}
        \sum_{n=1}^{N} p_{n}^{\ast}(Q_{C}^{\ast}) &= \sum_{n=1}^{N}\frac{(1+Q_{C}^{\ast})}{2b_{n}^{2}} + \sum_{n=1}^{N} \frac{r_{n}a_{n}}{2} - \nonumber \\
        &  \frac{(1+Q_{C}^{\ast})}{2}\left(1 + (1+Q_{C}^{\ast})\sum_{n=1}^{N} \frac{1}{b_{n}^{2}}\right) , \nonumber \\
        &= \sum_{n=1}^{N} \frac{r_{n}a_{n}}{2} - \frac{(1+Q_{C}^{\ast})}{2} . \label{eq:gamma_coop_pn}
    \end{align}
    Substituting (\ref{eq:gamma_coop_pn}) into (\ref{eq:gamma}), we obtain (\ref{eq:lemma_coop_gamma}). This completes the proof.
\end{proof}
\subsection{Proof of Theorem \ref{thm:NC}}
\begin{proof}
 From (\ref{eq:Un_2}), we obtain $\frac{\partial U_{n}}{\partial q_{n}}$ as
    \begin{equation}
        \frac{\partial U_{n}}{\partial q_{n}} = \frac{2r_{n}a_{n}}{\sqrt{(1+Q)^{2}+2b_{n}^{2}r_{n}a_{n}}+(1+Q)} - 1. \label{eq:Un_der2}
    \end{equation}
    From (\ref{eq:Un_der2}), we have
    \begin{align}
        \left.\frac{\partial U_{n}}{\partial q_{n}}\right \rvert_{Q=0}\leq 0 \implies \frac{2r_{n}a_{n}}{\sqrt{1+2b_{n}^{2}r_{n}a_{n}}+1} \leq 1 , \nonumber \\
        \implies \sqrt{1+2b_{n}^{2}r_{n}a_{n}} + 1 \geq 2r_{n}a_{n} \implies r_{n}a_{n} - \frac{b_{n}^{2}}{2} \leq 1 . \nonumber
    \end{align}
    Therefore, if $r_{n}a_{n}-\frac{b_{n}^{2}}{2} \leq 1$, we have $\left. \frac{\partial U_{n}}{\partial q_{n}}\right \rvert_{Q=0} \leq 0$.
    If $\max \limits_{n \in \mathcal{M}} \left(r_{n}a_{n}-\frac{b_{n}^{2}}{2}\right) \leq 1$, we have $r_{n}a_{n} - \frac{b_{n}^{2}}{2} \leq 1$ for all $n \in \mathcal{N}$.
    This in turn implies $\left.\frac{\partial U_{n}}{\partial q_{n}}\right \rvert_{Q=0} \leq 0$ for all $n \in \mathcal{N}$. From (\ref{eq:Un_der2}), we have
    \begin{equation}
        \left.\frac{\partial U_{n}}{\partial q_{n}} \right \rvert_{\{q_{n}=0, Q_{-n}>0\}} < \left.\frac{\partial U_{n}}{\partial q_{n}}\right \rvert_{\{q_{n}=Q_{-n}=0\}} \leq 0. \label{eq:NC_case1}
    \end{equation}
    It can be concluded from (\ref{eq:NC_case1}) that $q_{n}^{\ast}=0$ for any value of $Q_{-n}$ for all $n$. 

    We now consider the case when $\max \limits_{n \in \mathcal{N}} \left(r_{n}a_{n} - \frac{b_{n}^{2}}{2}\right) > 1$. 
    Since $\max \limits_{n \in \mathcal{N}} \left(r_{n}a_{n} - \frac{b_{n}^{2}}{2}\right) > 1$, $\left. \frac{\partial U_{n}}{\partial q_{n}} \right \rvert_{Q=0} >1$ holds for at least one CP, and hence, some CPs have an incentive to invest in the public investment. 
    Let us define $1+Q_{N}^{\ast} = \max \limits_{n \in \mathcal{N}} \left(r_{n}a_{n} - \frac{b_{n}^{2}}{2} \right)$. Note that $Q_{N}^{\ast}>0$ follows from $\max \limits_{n \in \mathcal{N}} \left(r_{n}a_{n} - \frac{b_{n}^{2}}{2}\right) > 1$. Further, $\left. \frac{\partial U_{n}}{\partial q_{n}} \right \rvert_{Q=Q_{N}^{\ast}} = 0$ for all $n \in \mathcal{M}$.
    Let $\{\hat{q}_{n}\}_{n \in \mathcal{N}}$ such that $\hat{q}_{n} \geq 0$ for all $n \in \mathcal{N}$ be any action profile of the CPs. Further, let $1+\mu = \sum_{n \in \mathcal{N}} \hat{q}_{N}$. We now consider the following cases. If $1+\mu > 1+Q_{N}^{\ast}$, $\left. \frac{\partial U_{n}}{\partial q_{n}}\right \rvert_{Q=\mu} < 0$ for all $n \in \mathcal{M}$. Thus, some CPs have an incentive to decrease their public investment. This implies that $1+\mu$ is not a Nash Equilibrium. Similarly, if $1+\mu < 1+Q_{N}^{\ast}$, $\left. \frac{\partial U_{n}}{\partial q_{n}}\right \rvert_{Q=\mu} > 0$ for all $n \in \mathcal{M}$. Therefore, some CPs have an incentive to increase their public investment. Thus, it is not a Nash Equilibrium. Lastly, if $1+\mu = 1+Q_{N}^{\ast}$, $\left. \frac{\partial U_{n}}{\partial q_{n}}\right \rvert_{Q=\mu} = 0$ for all $n \in \mathcal{M}$. It is a Nash Equilibrium if $\hat{q}_{n}=0$ for all $n \in \mathcal{N} \setminus \mathcal{M}$. This completes the proof.
\end{proof}
\subsection{Proof of Theorem \ref{thm:PoA}}
\begin{proof}
   We first consider the case when $\max\limits_{n \in \mathcal{N}} \left(r_{n}a_{n} - \frac{b_{n}^{2}}{2}\right) \leq 1$. This implies $\left(r_{n}a_{n} - \frac{b_{n}^{2}}{2}\right) \leq 1$ for all $n \in \mathcal{N}$. From Theorem \ref{thm:NC}, this in turn implies that $Q_{N}^{\ast}=0$. It can be verified that $\frac{\sqrt{1+2b_{n}^{2}r_{n}a_{n}}-1}{b_{n}^{2}} \leq 1$ follows from $r_{n}a_{n} - \frac{b_{n}^{2}}{2} \leq 1$ for all $n \in \mathcal{N}$. This leads to two possibilities which are as follows.
   \begin{itemize}
       \item $\sum_{n=1}^{N} \frac{\sqrt{1+2b_{n}^{2}r_{n}a_{n}}-1}{b_{n}^{2}} \leq 1$: In this case, from Lemma \ref{lem:lemma_coop}, we have $Q_{C}^{\ast}=0$, and hence, from (\ref{eq:eta}), it is clear that $\eta$ is unbounded.
       \item $\sum_{n=1}^{N} \frac{\sqrt{1+2b_{n}^{2}r_{n}a_{n}}-1}{b_{n}^{2}} > 1$: In this case, from Lemma \ref{lem:lemma_coop}, we have $Q_{C}^{\ast}>0$. However, since $Q_{N}^{\ast}=0$, from (\ref{eq:eta}), it follows that $\eta$ is unbounded.
   \end{itemize}
   Since $Q_{N}^{\ast}=0$ when $\max\limits_{n \in \mathcal{N}} \left(r_{n}a_{n} - \frac{b_{n}^{2}}{2}\right) \leq 1$, from (\ref{eq:gamma}), we have $\gamma_{N}=0$.

   Now we consider that $\max\limits_{n \in \mathcal{N}}\left(r_{n}a_{n} - \frac{b_{n}^{2}}{2}\right) > 1$. 
   From Theorem \ref{thm:NC}, $\left.\frac{\partial U_{m}}{\partial q_{m}}\right \rvert_{q_{m}=Q_{N}^{\ast}} = 0$ follows for each $m \in \mathcal{M}$, where $\mathcal{M}$ is as defined in (\ref{eq:set_M}). This implies 
   \begin{equation}
       \frac{\sqrt{(1+Q_{N}^{\ast})^{2}+2b_{m}^{2}r_{m}a_{m}}-(1+Q_{N}^{\ast})}{b_{m}^{2}}=1 ~\forall m \in \mathcal{M}. \label{eq:PoA_pf1}
   \end{equation}
   Using (\ref{eq:PoA_pf1}), we have
   \begin{align}
       \left. \frac{\partial U}{\partial Q}\right \rvert_{Q=Q_{N}^{\ast}} &= \sum_{n \in \mathcal{N} \setminus \mathcal{M}} \frac{\sqrt{(1+Q_{N}^{\ast})^{2}+2b_{n}^{2}r_{n}a_{n}}-(1+Q_{N}^{\ast})}{b_{n}^{2}} \nonumber \\
       &+ \sum_{n \in \mathcal{M}} \frac{\sqrt{(1+Q_{N}^{\ast})^{2}+2b_{n}^{2}r_{n}a_{n}}-(1+Q_{N}^{\ast})}{b_{n}^{2}} - 1, \nonumber
   \end{align}
    Thus, $\left. \frac{\partial U}{\partial Q}\right \rvert_{Q=Q_{N}^{\ast}}>0$. This implies $Q_{C}^{\ast}>Q_{N}^{\ast}$ as $\left. \frac{\partial U}{\partial Q}\right \rvert_{Q=Q_{C}^{\ast}} = 0$. This follows from the concavity of the utility function. Therefore, from (\ref{eq:eta}), $\eta>1$. Using (\ref{eq:opti_pn}), $p_{n}^{\ast}(Q_{N}^{\ast})$ for all $n \in \mathcal{N} \setminus \mathcal{M}$ is given by 
    \begin{equation}
        p_{n}^{\ast}(Q_{N}^{\ast}) = \left(\frac{\sqrt{(1+Q_{N}^{\ast})^{2}+2b_{n}^{2}r_{n}a_{n}}-(1+Q_{N}^{\ast})}{2b_{n}}\right)^{2}. \nonumber
    \end{equation}
    For all $n \in \mathcal{M}$, $p_{n}^{\ast}(Q_{N}^{\ast})$ simplifies to
    \begin{align}
        p_{n}^{\ast}(Q_{N}^{\ast}) &= \left(\frac{\sqrt{\left(r_{n}a_{n} - \frac{b_{n}^{2}}{2}\right)+2b_{n}^{2}r_{n}a_{n}} - \left(r_{n}a_{n}-\frac{b_{n}^{2}}{2}\right)}{2b_{n}}\right)^{2} , \nonumber \\
        &= \frac{b_{n}^{2}}{4} . \nonumber 
    \end{align}
    Therefore, $\gamma_{N}$ is given by
    \begin{equation}
        \gamma_{N} = \frac{Q_{N}^{\ast}}{\sum \limits_{n \in \mathcal{M}} \frac{b_{n}^{2}}{4} + \sum \limits_{n \in \mathcal{N} \setminus \mathcal{M}} \left(\frac{\sqrt{(1+Q_{N}^{\ast})^{2}+2b_{n}^{2}r_{n}a_{n}}-(1+Q_{N}^{\ast})}{2b_{n}}\right)^{2}} . \label{eq:gamma_n2}
    \end{equation}
    Clearly, (\ref{eq:gamma_n}) follows from (\ref{eq:gamma_n2}) for the special case of $|\mathcal{M}|=|\mathcal{N}|$. This completes the proof.
\end{proof}
\end{document}